\documentclass[twocolumn,showpacs,preprintnumbers]{revtex4}
\usepackage{amsmath,amssymb}
\usepackage{graphicx}
\usepackage{dcolumn}
\usepackage{bm}

\begin{document}

\title{Entanglement-preserving frequency conversion in cold atoms}
\author{A.Gogyan}
\author{Yu.Malakyan}
\email{yumal@ipr.sci.am}

\affiliation{Institute for Physical Research, Armenian National Academy of Sciences,
Ashtarak-2, 378410, Armenia }
\date{\today }

\begin{abstract}
We propose a method that enables efficient frequency conversion of quantum information based on recently demonstrated strong parametric coupling between two single-photon pulses propagating in a slow-light atomic medium at different group velocities. We show that an incoming single-photon state is efficiently converted into another optical mode in a lossless and shape-conserving manner. The persistence of initial quantum coherence and entanglement within frequency conversion is also demonstrated. We first illustrate this result for the case of small frequency difference of converted photons, and then discuss the modified scheme for conversion of photon wavelengths in different spectral ranges. Finally we analyze the generation of a narrow-band single-photon frequency-entangled state.
\end{abstract}

\pacs{42.50.Dv, 42.50.Gy, 42.65.Ky, 03.67.-a} \maketitle




\section{\protect\normalsize INTRODUCTION}
Frequency conversion of quantum states of light and redistribution
of optical information between different quantum fields is an
important and desirable tool for interfacing of quantum
communication lines with photon memory units. Quantum frequency
conversion (QFC) with preserving a quantum state has been hitherto
realized in the nonlinear crystals using the process of parametric
up-conversion \cite{huang, giorgi, kwiat, albota, tanzilli}. For
implementation of QFC  in atomic ensembles, a technique for light
storage and its subsequent retrieval \cite{fleis} at another
optical frequency under the conditions of electromagnetically
induced transparency (EIT) \cite{harris} was employed. However,
despite the success of proof-of-principle experiments performed in
a four-level double $\Lambda$-type atomic medium
\cite{zibrov,wang}, to date no true demonstration of
information-preserving frequency conversion has been given. The
major difficulties inherent to these schemes are unavoidable
losses and shape distortion of a weak (quantum) light pulse during
its storage and retrieval by means of EIT.

In this paper, we propose a protocol for QFC  in atomic ensembles
free from the above drawbacks. Our method makes use of recently
demonstrated \cite {sis} efficient parametric coupling between two
single-photon pulses propagating in a slow-light medium.  The
latter is an ensemble of atoms which interact with two quantum
fields in a V-type configuration, while the upper electric-dipole
forbidden transition is driven by a classical and constant field
inducing a magnetic dipole or an electric quadrupole transition
between the two upper levels (Fig.1). The role of classical field
is twofold. First, it creates parametric coupling between the
photons and, second, it induces medium transparency for the
intensity values where the EIT conditions are fulfilled for both
quantum fields. It is clear, however, that implementation of the
medium transparency leads evidently to degradation of parametric
interaction between the fields. Nevertheless, we demonstrate below
how an incoming signal-photon
\begin{figure}[b] \rotatebox{0}{\includegraphics*
[scale = 0.5]{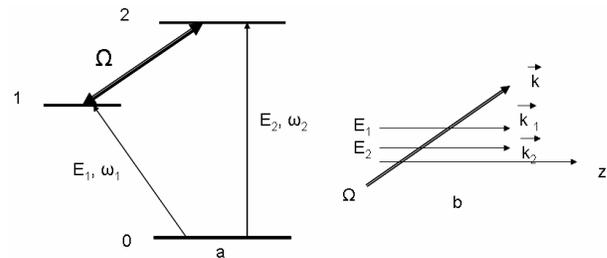}} \caption{(a) Level scheme of atoms
interacting with quantum fields $E_{1,2}$ and classical driving
field of Rabi frequency $\Omega$. (b) Geometry of fields
propagation.}
\end{figure}
state is efficiently converted under EIT conditions into second
optical mode in a lossless manner conserving at the same time the
pulse shape and initial entanglement. The easiest way to realize
this is obviously the case of equal group velocities of quantum
fields, which, however, is almost never met in practice (see
below). Meanwhile, for realistic atomic systems, we predict the
conversion efficiency as high as $\sim90\%$. In essence, while the
process involves only one photon at a time, this system allows for
the realization of strong interaction of individual photons with
each other, even if the two pulses propagate in the medium with
different group velocities. Our approach offers another important
advantage: unlike the previous proposals for QFC based on the
light storage and its retrieval, we apply here only one and,
moreover, constant driving field for control the conversion
efficiency that makes our method much favorable for future
applications. One obvious limitation of considered process is that
it is effective in a relatively narrow frequency range associated
with specific atoms. We note, however, that recently a source of
narrow-band, frequency tunable single photons with properties
allowing exciting the narrow atomic resonances has been created
\cite{chou, eis, mats, yuan}.

In next section we briefly describe a three-level model for
parametric interaction between two quantum fields and give the
analytic solution for the field operators obtained in \cite {sis}.
Then, in section III we calculate the intensities of quantum
fields and find the conversion efficiency, as well as discuss  the
ability of the system to convert coherently the photons with
wavelengths in different spectral ranges. In section IV we analyze
 the quantum properties of frequency conversion process
and show how a partial conversion leads to generation of
frequency entangled single-photon state. Finally, in
section IV we summarize our results.

\section{\protect\normalsize  PARAMETRIC INTERACTION BETWEEN TWO SINGLE-PHOTON PULSES}

We outline here an approach that allows the propagation dynamics
to be solved exactly, while the detailed description of the model
can be found in our previous paper \cite {sis}. We consider an
ensemble of cold atoms with level configuration depicted in Fig.1.
Two quantum fields
\begin{equation*}
E_{1,2}(z,t)=\sqrt{\frac{\hbar \omega _{1,2}}{2{\varepsilon }_{0}V}}%
\hat{\mathcal E}_{1,2}(z,t)\exp [i(k_{1,2}z-\omega _{1,2}t)]+h.c.
\end{equation*}
co-propagate along the $z$ axis and interact with the atoms on the
transitions $0 \rightarrow 1$ and $0 \rightarrow 2$, respectively,
while the upper electric-dipole forbidden transition $1
\rightarrow 2$ is driven by a classical and constant field with real
Rabi frequency $\Omega $, and $V$ is the quantization volume taken
to be equal to interaction volume. The electric fields are
expressed in terms of the operators $\hat{\mathcal E}_{i}(z,t)$,
and the medium is described using atomic
operators $\hat{\sigma}_{\alpha \beta }(z,t)$ $=\frac{1}{N_{z}}\underset{i=1}%
{\overset{N_{z}}{\sum }}\mid \alpha $ $\rangle _{i}\langle \beta
\mid $ averaged over the volume containing many atoms
$N_{z}=\frac{N}{L}dz\gg 1$ around position $z$, where $N$ is the
total number of atoms and $L$ is the length of the medium. In the
rotating wave picture the interaction Hamiltonian is given by
\begin{equation*}
H=-\hbar \frac{N}{L}\int\limits_{0}^{L}dz[g_{1}\hat{\mathcal E}_{1}\hat{\sigma}%
_{10}e^{ik_{1}z}+g_{2}\hat{\mathcal E}_{2}\hat{\sigma}_{20}e^{ik_{2}z}+\Omega \hat{%
\sigma}_{21}e^{ik_{\parallel }z}
\end{equation*}
\begin{equation}
+h.c.]
\end{equation}
Here $k_{\parallel }=\vec{k}_{d}\hat{e}_{z}$ \ is the projection
of the wave-vector of the driving field on the $z$ axis,
$g_{\alpha }=\mu _{\alpha 0}\sqrt{\omega _{i}/(2\hbar {\varepsilon
}_{0}V)}$ is the atom-field coupling constants with $\mu _{\alpha
\beta }$ being the dipole matrix element of the atomic transition
$\alpha \rightarrow \beta $. We assume that the process is running at
low temperature in order to avoid the Doppler broadening, which in
a cold atomic sample is smaller than all relaxation rates, and
consider the case of exactly resonant interaction with all fields.

To implement our QFC scheme in a dense atomic medium, several
conditions must be satisfied.
 First, the photon absorption must be strongly reduced
\begin{equation}
\kappa_{i}L\ll 1
\end{equation}
where $\kappa_{i}={g_{i}^{2}\Gamma_{j} N}/{c\Omega ^{2}}, i,j=1,2$
and $i\neq j$, are the field absorption coefficients,  and $\Gamma
_{1,2}$ are the optical transverse relaxation rates involving,
apart from natural decay rates $\gamma _{1,2}$ of the excited
states 1 and 2, the dephasing rates in corresponding transitions.
The latter are caused by atomic collisions and escape of atoms
from the laser beam. However, in the ensemble of cold atoms the
both effects are negligibly small compared to $\gamma _{1,2}$, so
that $\Gamma _{1,2}$=$\gamma_{1,2}/2$. We emphasize that  $\gamma_{i}$
is a sum of the partial decay rate of $i$-th upper level to the ground state
0 and the rate of population leak from the $i$-th level towards the states outside of the system.
The limit of Eq.(2) is readily achieved, if the condition of electromagnetically induced
transparency (EIT, ref. \cite {harris}) $\Omega \gg \Gamma _{1,2}$
is satisfied for both transitions coupled to the weak-fields. Note
that the three-level configurations 0-2-1 and 0-1-2 form the
$\Lambda$- and ladder EIT-systems with the decoherence time
$\Gamma_{1}^{-1}$ and $\Gamma_{2}^{-1}$, respectively. Second, the
initial spectrum of quantum fields should be contained within the
EIT window $\Delta \omega _{EIT}={\Omega ^{2}}/({\Gamma
\sqrt{\alpha }})$ \cite{fleis}, resulting in little pulse
distortion from absorption, i.e.
\begin{equation}
\Delta \omega _{EIT}T\geq 1
\end{equation}
where $T$ is the initial pulse width, $\alpha =\mathcal{N}\sigma
L$ is the optical depth, $\sigma =\frac{3}{4\pi }\lambda ^{2}$ is
the resonant absorption cross-section, and $\mathcal{N}$ is the
atomic number density. It is also desirable that the pulse
broadening should be minimal. The source of the latter is the
various orders of dispersion, beginning from the group-velocity
dispersion proportional to the second time derivative of the
fields $\frac{\partial^{2} }{\partial t^{2}}\hat{\mathcal
E}_{i}(z,t)$. The change in pulse width due to this term after
propagating through the medium can be roughly estimated by
treating the weak fields for a time classically and ignoring the
parametric contribution to broadening.  Then, for an initial
Gaussian pulse $\mathcal E_{1}(0,t)=\mathcal
E_{0}exp(-\frac{2t^{2}}{T^{2}})$, the simple calculations yield
\begin{equation*}
T_{i}(L)=T\sqrt{1+\frac{16L}{v_{i}T^{2}\Omega}}
\end{equation*}
with $ v_{i}={c\Omega ^{2}}/{g_{i}^{2}N}\ll c$ being the pulse
group velocity. It is seen that the spreading of the quantum
pulses caused by the group-velocity dispersion can be neglected,
if
\begin{equation}
\frac{16\tau_{i}}{T^{2}\Omega}\leq 1
\end{equation}
where $\tau_{i}=L/v_{i}$ is the pulse propagation time and, in
fact, the pulse delay relative to the pulse which travels the
same distance in vacuum. It is worth noting that upon satisfying
the condition (4), the pulse broadening due to next orders of
dispersion is more suppressed.

In \cite {sis} we have shown that under the conditions (2-4) the
propagation equations for the quantum field operators, in the
slowly varying envelope approximation, take the simple form :
\begin{equation}
\left( \frac{\partial }{\partial z}+\frac{1}{v_{1}}\frac{\partial
}{\partial t}\right) \hat{\mathcal E}_{1}(z,t)=-i\beta
\hat{\mathcal E}_{2}
\end{equation}%
\begin{equation}
\left( \frac{\partial }{\partial z}+\frac{1}{v_{2}}\frac{\partial
}{\partial t}\right) \hat{\mathcal E}_{2}(z,t)=-i\beta
\hat{\mathcal E}_{1}
\end{equation}
where $\beta ={g_{1}g_{2}N/c\Omega }$ is the coupling constant of
parametric interaction between the quantum  fields and decreases
with increase of coupling field intensity.

The formal solution of Eqs.(6,7) in the region $%
0\leqslant z\leqslant L$ is given by
\begin{equation*}
\hat{\mathcal E}_{i}(z,t)=\hat{\mathcal E}_{i}(0,\ t-z/v_{i})+\int\limits_{0}^{z}dx\{\hat{\mathcal E}%
_{i}(0,\ t-z/v_{j}-\frac{\Delta v_{ji}}{v_{i}v_{j}}x)
\end{equation*}%
\begin{equation}
\times \frac{\partial J_{0}(\psi )}{\partial z}-i\beta \
\hat{\mathcal E}_{j}(0,\ t-z/v_{i}-\frac{\Delta
v_{ij}}{v_{i}v_{j}}x)\ J_{0}(\psi )\}
\end{equation}%
where $i,j=1,2$ \ and $j\neq i.$ The Bessel function $\ J_{0}(\psi
)$ depends on $z$ via $\psi =2\beta \sqrt{x(z-x)}$ , $\Delta
v_{ij}=v_{i}-v_{j}$ is the difference of group velocities. In
deriving this solution we have assumed that the phase-matching
condition $\Delta k=k_{2}-k_{1}-k_{\parallel }=0$ is fulfilled in
the medium.

In the limit of equal group velocities $v_{2}=v_{1}=v$ the Eqs.
(7) are reduced to a simple form
\begin{equation}
\hat{\mathcal E}_{i}(z,t)=\hat{\mathcal E}_{i}(0,\tau )cos(\beta
z)-i\hat{\mathcal E}_{j}(0,\tau )sin(\beta z)
\end{equation}
where $\tau =t-z/v$, $j\neq i$. Note, however, that this case is
practically never realized, since the atom-field coupling
constants $g_{1}$ and $g_{2}$ are defined by the partial decay
rates of upper atomic 1 and 2 levels into the ground state $0$, which
are different even if the total decay rates are the same $\gamma
_{1}=\gamma _{2}$, as is the case, for example, in a V-type atoms
with hyperfine-level structure.

\section{\protect\normalsize QUANTUM EFFICIENCY OF CONVERSION}

Our aim is to show that the proposed scheme is suitable for
converting individual photons at one frequency to another
frequency while preserving initial quantum coherence that results
in a generation of frequency-entangled single-photon state. To
this end we analyze the evolution of the input state $\mid \psi
_{in}\rangle = \mid 1_{1}\rangle \otimes \mid 0_{2}\rangle $
consisting of a single-photon wave packet at $\omega _{1}$
frequency, while $\omega _{2}$ field is in the vacuum state. The
similar results are clearly obtained in the case of one input
photon at $\omega _{2}$ frequency. We assume that initially the
$\omega _{1}$ pulse is localized around $z=0$ with a given
temporal profile $f_{1}(t)$:
\begin{equation}
\langle 0\mid \hat{\mathcal E}_{1}(0,t)\mid \psi _{in}\rangle =\langle 0\mid \hat{\mathcal E}%
_{1}(0,t)\mid 1_{1}\rangle =f_{1}(t)
\end{equation}%
In free space, $\hat{\mathcal E}_{1}(z,t)=\hat{\mathcal
E}_{1}(0,t-z/c)$ and we have
\begin{equation}
\langle 0\mid \hat{\mathcal E}_{1}(0,t-z/c)\mid 1_{1}\rangle
=f_{1}(t-z/c).
\end{equation}%
The intensities of the fields at any distance in the region $%
0\leqslant z\leqslant L$ are given by%
\begin{equation}
\langle I_{i}(z,t)\rangle =\mid \langle 0\mid \hat{\mathcal
E}_{i}(z,t)\mid \psi _{in}\rangle \mid ^{2}
\end{equation}
The quantum efficiency (QE) is determined as the ratio
$n_{2}(L)/n_{1}(0)$, where $n_{i}(z)=\langle \psi_{in}\mid \hat
n_{i}(z)\mid \psi _{in}\rangle$ are the mean photon numbers and
$\hat n_{i}(z)$ are the dimensionless operators for number of
photons that pass each point on the z axis in the whole time
\begin{equation}
\hat n_{i}(z)=\frac{c}{L}\int dt\hat{\mathcal
E}_{i}^{+}(z,t)\hat{\mathcal E}_{i}(z,t)
\end{equation}%
\begin{figure}[b]
\rotatebox{0}{\includegraphics* [scale = 0.9]{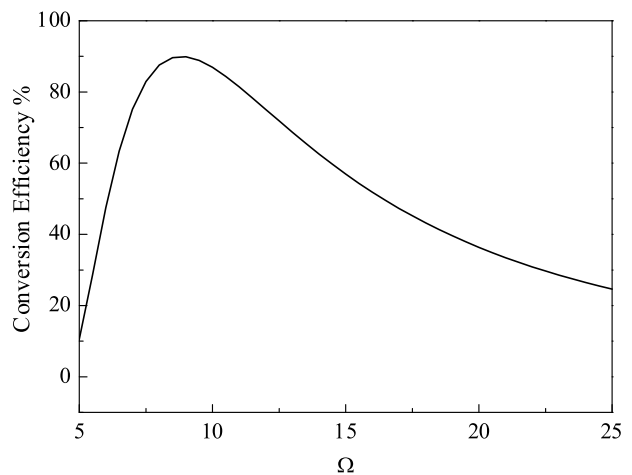}} \caption{
Quantum efficiency of optical field conversion from $\lambda_{1}=795 nm$ into  $\lambda_{2}=780 nm$ in $^{87}$Rb as a function of driving field Rabi frequency $\Omega$ (given in units of $\Gamma_{2}$) for the medium length $L=100\mu m$. For the rest of parameters see the text.}
\end{figure}
Using Eqs.(7, 9-12) and recalling that $\langle 0\mid
\hat{\mathcal E}_{2}(0,t)\mid \psi _{in}\rangle =0$, we calculate
$n_{i}$ numerically  and show in Fig.2 the QE as a function of
driving field Rabi frequency $\Omega$ in the case of a Gaussian
input pulse $f(t)=C\exp [-2t^{2}/T^{2}]$, which is normalized as
$\frac{c}{L}\int\mid f_{1}(t)\mid ^{2}dt=1$ indicating that the
number of impinged photons is one. The atomic sample is chosen to
be $^{87}Rb$ vapor with the ground state $5S_{1/2}(F_{g}=2)$ and
exited states $5P_{1/2}(F_{e}=1)$, $5P_{3/2}(F_{e}=1)$  as the
atomic states $0$ and  $1,2$ in Fig.1, respectively. For
calculations we used the following parameters: light wavelength of
quantum fields $\lambda_{1}$ =795 nm and $\lambda_{2}$ =780 nm,
$\Gamma=\Gamma_{2} =2\Gamma_{1}=2\pi \times 3$ MHz, atomic density
$\mathcal{N} \sim 10^{13}$cm$^{-3}$ in a trap of length $L\sim
100$ $\mu $m, and the input pulse duration $T\simeq 20$ ns, for
which we have $v_{1}\sim 1.25\cdot10^{4}m/s$, $v_{2}\sim
0.5v_{1},$ and $\kappa_{i}L< 0.1$. All of these parameters appear
to be within experimental reach, including the initial
single-photon wave packets with a pulse length of several tens of
nanoseconds \cite{mats, yuan}, as well as high power narrow-band
cw THz radiation at a frequency $\sim 7$ THz \cite {hubers}
resonant to the fine splitting $\omega(5P_{3/2}-5P_{1/2})$ in
rubidium.

It is worth noting that in most cases of alkali atoms the spectral
separation between $D2$ and $D1$ lines lies in the operation range
of tunable terahertz silicon and quantum cascade lasers \cite
{barkan, pavlov} that may give novel important applications of THz
radiation such as implementation of QFC in atomic ensembles.

\begin{figure}[b] \rotatebox{0}{\includegraphics* [scale =
0.9]{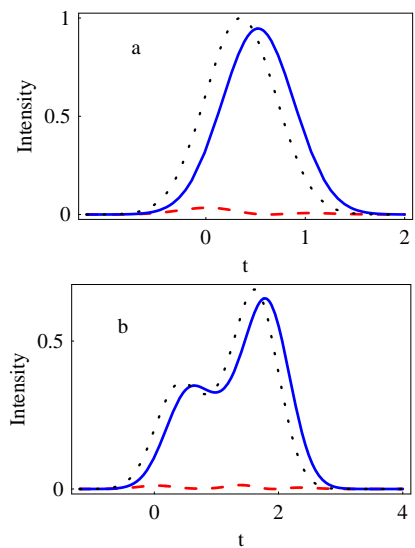}} \caption{(Color online) Output shapes of $\lambda_{1}$ (dashed line) and $\lambda_{2}$ (solid line) modes in the cases of
Gaussian (a) and double-hump (b) input $\lambda_{1}$ pulses for $\Omega = 8\Gamma$ and other parameters as in Fig.2. The dotted line displays the $\lambda _{1}$ pulse propagating in the medium with  $\beta =0$. The time is given in units of $T$.}
\end{figure}
In Fig.2 we do not show the region of small values of $\Omega$,
where the solution Eq.(7) is not valid. It is seen that the QE
reaches its maximum $\sim90\%$ at $\Omega\sim 8 \Gamma$, which is
very close to the corresponding value of $\Omega$ following from
Eq.(8) for equal group velocities. To demonstrate the shape-conserving
character of frequency conversion in our scheme we present in Fig.3 the
results for two different shapes of initial $\omega_{1}$ pulse. Fig.3a shows the
case of Gaussian shaped input pulse having a duration T, while in Fig.3b
the latter is taken as two superimposed Gaussian pulses of the same
duration T. It is seen that in both cases the generated $\omega_{2}$ mode
reproduces almost identically the initial form of the input $\omega_{1}$ pulse,
which is evidently a direct consequence of linear relationship
between quantum fields in Eqs.(5) and (6).

At other values of $\Omega$ the incoming $\omega _{1}$ field is
converted into $\omega _{2}$ mode not completely, but different
amount of mode conversion is attainable, thus enabling
redistribution of quantum information between the two quantum fields.
By adjusting the driving field intensity, a single-photon state
entangled over the two optical modes $\omega _{1}$ and $\omega _{2}$
with the given intensities can be generated. In particular, such a
state with equal intensities of the modes is created for two
values of $\Omega\sim 6\Gamma$ and $\Omega\sim 18\Gamma$ (Fig.4),
which, however, are different in their physical content. The
higher values of $\Omega$ correspond to large group velocities of
the pulses, while the parametric coupling is reduced. As a result,
during propagation in the medium the $\omega _{1}$ field has time
to be converted into $\omega _{2}$ mode only partially and, owing
to $v_{2}<v_{1}$, the $\omega _{2}$ pulse is slightly delayed with
respect to the signal $\omega _{1}$ pulse that is just observed in
Fig.4 (left column). On the contrary, at small values of $\Omega$
the pulses travel slower, while the parametric coupling is
enhanced. Consequently, first the $\omega _{1}$ field is almost
completely converted into the $\omega _{2}$ pulse at small
distances (nearly 60 $\mu m$), and then the latter is partially
transformed back into the $\omega _{1}$ photon at the end (100$\mu
m$) of the medium. This time the newly generated $\omega _{1}$
pulse is clearly behind the $\omega _{2}$ pulse (Fig.4, right
column) with a delay which is longer than in the previous case. By
these any input optical mode is split into two optical modes of
different frequencies in a controllable way preserving at the same
time the total number of photons, which  is determined in each
mode by the areas of the corresponding peaks. In the absence of
losses, the conservation law for photon numbers follows from
Eqs.(5) and (6) as
\begin{equation}
\frac{\partial }{\partial z}(n_{1}(z)+n_{2}(z)) =0
\end{equation}%
with taking into account that $\langle 0\mid \hat{\mathcal
E}_{i}(z,t\rightarrow\pm\infty)\mid \psi _{in}\rangle=0$. At the
input of the medium  $n_{1}(0)$=1 and $n_{2}(0)$=0, so that from
Eq.(13) one has $n_{1}(z)+n_{2}(z)$=1.

\begin{figure}[b]
\rotatebox{0}{\includegraphics* [scale = 0.65]{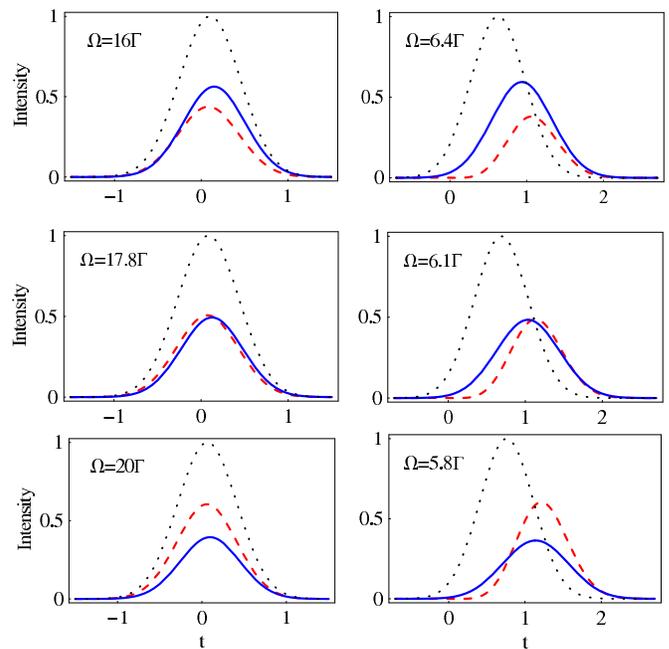}}
\caption{(Color online) The same as in Fig.3a, but for different values of $\Omega$.}
\end{figure}

So far we considered the pulses with a duration comparable to the
lifetime of atomic 1 and 2 levels. However, the conditions (2-4)
do not limit critically the pulse length provided that the
spectral bandwidth of quantum fields is still negligible compared
to the separation between the upper atomic levels. These
conditions can be easily satisfied also for ultrashort pulses, if
low atomic densities (lower by several orders of magnitude) are
used \cite {remind}, thus making our scheme able for efficient
frequency conversion in this case as well, which is a permanent
need in quantum communication tasks.

Thus, by combining the clear properties of parametric
interaction between the photons with that the converted light
is within the narrow atomic resonant width, one can
significantly expand the ability to distribute the quantum states
of narrowband spectra. Above we have demonstrated this possibility
in the case of small frequency difference of converted photons that
can be used, in particular, for addressable excitation of registers
of quantum computers, which may be trapped ions or atoms, atomic
ensembles, quantum dots etc. At the same time, there is considerable
interest in exploring similar mechanisms that enable coherent
conversion of photons with much larger difference in wavelengths.
Such a mechanism would enable quantum information to be passed over
long distances by telecom fibers, which impose the infrared
(IR) wavelengths, and then to be mapped into the states of atoms,
which interact resonantly with quantum fields at visible frequencies.
Within our approach, the required frequency conversion is easily
realized by slightly modifying the interaction configuration in Fig.1
and applying a second classical field on the transition $0\rightarrow 3$ with
the Rabi frequency $\Omega_{0}$ (Fig.5a). The details of this analysis will
be published elsewhere. Here we only outline how the same high
conversion efficiency as in the previous case is obtained, but
now for $\omega_{1}$  and $\omega_{2}$ light modes, which are in the visible and
IR regions, respectively.
\begin{figure}[b]{\includegraphics*
[scale = 0.75]{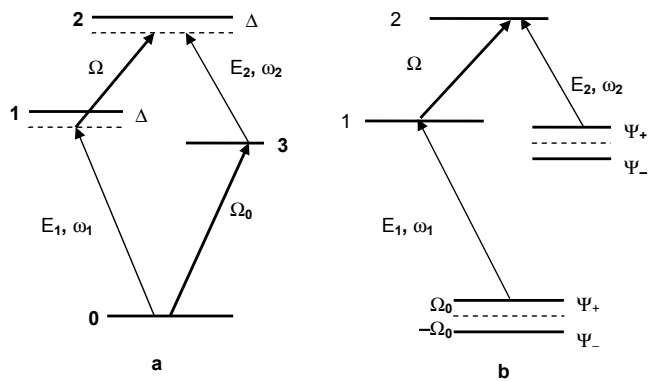}}
\caption{Modified scheme for conversion of quantum fields $E_{1}$ and $E_{2}$ with essentially different wavelengths in the basis of bare (a) and dressed (b) atomic states.}
\end{figure}
In Fig.5a the quantum fields $E_{1}$ and $E_{2}$
have equal one-photon detuning $\Delta$, while the driving fields  $\Omega$ and $\Omega_{0}$ are exactly resonant with the corresponding transitions. We suppose   $\Omega_{0}\gg \Gamma_{3}$, so that the bare atomic states 0 and 3 are
split into a doublet of dressed states $\Psi_{\pm}=( |3 \rangle \mp | 0\rangle )/ \sqrt 2$ , which
are well separated by 2$\Omega_{0}$. If now $\Delta = \Omega_{0}$, then taking into account
that in this case the excitation of the atoms from $\Psi_{-}$ can be neglected, since it is strongly suppressed by the factor of $( \Gamma_{3}/ \Omega_{0})^{2}$, the interaction scheme is reduced to Fig.5b, where  $\Psi_{+}$ serves as the ground state(correspondingly, for alternative choose $\Delta = -\Omega_{0}$, the ground state of modified system is $\Psi_{-}$ ), thus making this scheme identical to that of Fig.1. Under
the approximations of Eqs.(2,3,4), we obtain for quantum field
operators the same Maxwell equations (5,6) with new parametric
coupling constant $\beta =g_{1}g_{2}N/(4 c \Omega)$, where $4$ in denominator emerges from the fact that only a half of the atoms in ground state take part in the process and the atom-field coupling constants $g_{1}$ and $g_{2}$ decrease by factor $\sqrt 2$. A key feature of this model is that $g_{1}$ and $g_{2}$ are approximately equal. Note that this is not the case in Fig.1, if we take here $\omega_{1}$  and $\omega_{2}$ in  different regions of spectrum, and thereby the conditions (2-4) cannot be satisfied for both quantum fields $E_{1,2}(z,t)$ simultaneously. As an illustration to the modified model in Fig.5, we again consider the sample of $^{87}Rb$ vapor with the states $5S_{1/2},  5P_{3/2},  4D_{3/2}$,  and $5P_{1/2}$ being the atomic states 0, 1, 2 and 3 of Fig.5a, respectively. In this case the quantum field wavelengths are $\lambda_{1}\sim 780nm$ and $\lambda_{2}\sim$ 1,47$\mu m$ and $g_{2}/g_{1}\sim 0.96$ that provides high efficiency of IR field conversion into the visible domain and vice versa.

\section{\protect\normalsize GENERATION OF FREQUENCY ENTANGLED SINGLE-PHOTON STATE}

Now we discuss the quantum properties of output single-photon
state. We describe it in terms of quantized wave packets at the
frequencies $\omega_{1}$ and $\omega_{2}$ containing the mean
photon numbers $n_{i}(L)=\frac{c}{L}\int dt \mid
\Phi_{i}(L,t)\mid^2$, where $\Phi_{i}(L,t)= \langle 0\mid
\hat{\mathcal E}_{i}(L,t)\mid \psi_{in}\rangle$ are the wave
functions of the modes. Using Eq.(12) and the commutation
relations \cite {sis}
\begin{equation}
\lbrack \hat{\mathcal E}_{i}(z,t),\hat{\mathcal
E}_{j}^{+}(z,t^{\prime})]=\frac{L}{c}\delta _{ij}\delta
(t-t^{\prime})
\end{equation}
we obtain the output single-photon state as the eigenstate of
total photon number operator
\begin{equation*}
(\hat n_{1}(L)+\hat n_{2}(L))\mid \psi_{out}\rangle=\mid
\psi_{out}\rangle
\end{equation*}
which yields
\begin{equation}
\mid \psi_{out}\rangle=\frac{c}{L}\int dt
[\Phi_{1}(L,t)\hat{\mathcal
E}_{1}^{+}(L,t)+\Phi_{2}(L,t)\hat{\mathcal E}_{2}^{+}(L,t)]\mid
0\rangle
\end{equation}
Let us introduce the operators of creation of single-photon wave
packets at frequencies $\omega_{i}$ associated with mode functions
$\Phi_{i}(L,t)$, where $i$ labels the members of the denumerably
infinite set.  For $i=1,2$ they are given by
\begin{equation}
\hat c_{1,2}^{+}=N_{i}^{1/2}\int dt \Phi_{1,2}(L,t)\hat{\mathcal
E}_{1,2}^{+}(L,t)
\end{equation}
with the normalization constants
\begin{equation*}
N_{i}=\frac {c}{L}(\int dt \mid\Phi_{1}(L,t)\mid^2)^{-1}
\end{equation*}
These operators create the single-particle states in the usual way
by acting on the vacuum state $\mid 0\rangle$
\begin{equation}
\hat c_{i}^{+}\mid 0\rangle=\mid 1_{i}\rangle
\end{equation}
and have the standard boson commutation relations
\begin{equation}
[\hat c_{i},\hat c_{j}^{+}]=\delta_{ij}
\end{equation}
following from Eq.(14).

Note that this definition of quantized wave packets is only
useful, if the mode spectra are much narrower compared to the mode
spacing that has been suggested from the very beginning. Now, for
the algebra (18) we choose the representation of infinite product
of all vacua
\begin{equation}
\mid 0\rangle=\prod_{i}\mid 0_{i}\rangle=\mid 0_{1}\rangle\mid
0_{2}\rangle\prod_{i\neq 1,2} {\mid 0_{i}\rangle}.
\end{equation}
However, since in our problem we deal with two frequency modes,
while the other modes are not occupied by the photons and, hence,
are not taken into account during the measurements, the vacuum may
be reduced to $\mid 0\rangle=\mid 0_{1}\rangle\mid 0_{2}\rangle$.
Then the single-photon state (15) can be written as
\begin{equation*}
\mid \psi_{out}\rangle=r_{1}\hat c_{1}^{+}\mid 0_{1}\rangle\mid
0_{2}\rangle+ r_{2}\mid 0_{1}\rangle\hat c_{2}^{+}\mid
0_{2}\rangle=
\end{equation*}
\begin{equation}
=r_{1}\mid 1_{1}\rangle\mid 0_{2}\rangle+r_{2}\mid
0_{1}\rangle\mid 1_{2}\rangle
\end{equation}
with $r_{1,2}=\sqrt{n_{1,2}(L)}$.

Thus, in general case of $r_{1,2}\neq 0$, the system produces a
photonic qubit, i.e., a single-photon state entangled in two
distinct frequency modes with the known wave functions
$\Phi_{1,2}(L,t)$. The complete conversion of input $\omega_{1}$
mode depicted in Fig.3 corresponds to pure output state $\mid
\psi_{out}\rangle=\mid 0_{1}\rangle\otimes\mid 1_{2}\rangle$.

Finally, we demonstrate the persistence of initial quantum
coherence and entanglement within the QFC in a simple case of
input  single-photon state entangled in two well-separated
temporal modes or time-bins \cite {brend}(the concept of the single-photon entanglement is discussed in \cite{sis})
\begin{equation}
\mid \psi_{in} \rangle = (a \mid 1_{1}\rangle_{t} \mid
0_{1}\rangle_{t+\tau} +  b \mid 0_{1}\rangle_{t} \mid
1_{1}\rangle_{t+\tau})\otimes \mid 0_{2}\rangle
\end{equation}
where $\mid 0_{1}\rangle_{t}$ and $\mid 1_{1}\rangle_{t}$ denote
Fock states with zero and one $\omega_{1}$ photon, respectively,
at the time $t$ and $\mid a\mid ^{2} +\mid b \mid^{2}=1$. Suppose
that the single-photon wave packets $\mid 1_{1}\rangle_{t}$ and
$\mid 1_{1}\rangle_{t+\tau}$ are characterized by temporal
profiles $f_{0}(t)$ and $f_{\tau}(t)$, respectively, which are not
overlapped in time due to large time shift $\tau$. Then, using the
Eqs.(7) and (10), the wave function of output $\omega_{2}$ mode is
readily calculated to be
\begin{equation*}
\Phi_{2}(L,t)= \langle 0\mid \hat{\mathcal E}_{2}(L,t)\mid
\psi_{in}\rangle=a\Phi_{2,0}(L,t)+b\Phi_{2,\tau}(L,t)
\end{equation*}
where
\begin{equation*}
\Phi_{2,0(\tau)}(L,t)=-i\beta \int\limits_{0}^{L}dxf_{0(\tau)}(X)\
J_{0}(\psi )\,
\end{equation*}
with $X=t-L/v_{2}-x\Delta v_{21}/{v_{1}v_{2}}$. Consequently, from
Eq.(16) the creation operator $c_{2}^{+}$ can be represented as a
sum of creation operators of the two temporal modes at $\omega_{2}$
frequency. Following the procedure discussed above the output
state in the case of complete conversion ($r_{1}=0$) is eventually
found in the form
\begin{equation}
\mid \psi_{out} \rangle = \mid 0_{1}\rangle\otimes(a \mid
1_{2}\rangle_{t} \mid 0_{2}\rangle_{t+\tau} +  b \mid
0_{2}\rangle_{t} \mid 1_{2}\rangle_{t+\tau})
\end{equation}
showing that the initial $\omega_{1}$ qubit is transformed into
another at $\omega_{2}$ frequency with the same complex amplitudes
$a$ and $b$, thus preserving the original amount of entanglement.
Two different methods are available for experimental test of
coherence transfer during the frequency conversion. In the first case,
following the work \cite{giorgi} the single-photon interference for incoming
and outgoing photons can be measured and compared to each other,
thus revealing an equal phase periodicity in the two interference
patterns in agreement with Eqs.(21) and (22). The second method \cite{huang,tanzilli}
is based on photon-pair interference; the input $\omega_{1}$ photon is first
nonclassically correlated with another photon at $\omega_{3}$, which does not
take part in the frequency conversion process. Then, after complete
conversion of the $\omega_{1}$ photon into  $\omega_{2}$ photon, the entanglement will
arise between the $\omega_{2}$ and  $\omega_{3}$ photons and, hence, the strong two-photon
interference between the latter will demonstrate the transfer of input
qubit at $\omega_{1}$ frequency into another at $\omega_{2}\neq \omega_{1}$ frequency. To realize
this program for narrowband fields, as is the case here, we propose
to employ the Duan, Lukin, Cirac, and Zoller (DLCZ) protocol \cite{DLCZ,sis1}
for generation of nonclassically correlated pair of Stokes  $\omega_{1}$- and
anti-Stokes  $\omega_{3}$-photons. Note that the key features of DLCZ protocol
have been confirmed in many experiments, including strong quantum
correlations between $\omega_{1}$ and $\omega_{3}$ fields \cite{kuzm}.

\section{\protect\normalsize CONCLUSIONS}

Summarizing, we have proposed and analyzed a simple scheme of
parametric frequency conversion of optical quantum information in
atomic ensembles. We have demonstrated remarkable properties
of this scheme such as minimal loss and distortion of the pulse
shape and the persistence of initial quantum coherence and
entanglement that make it superior against the previous schemes of
the QFC in atomic media based on release of stored light state
via EIT. Moreover, the narrowband property of single photons and
high efficiency of entanglement generation profit the present
mechanism to serve as an ideal candidate for frequency conversion
and redistribution of optical information in quantum information
processing and quantum networking.

\bigskip

This work was supported by the ISTC Grant No.A-1095 and INTAS
Project Ref.Nr 06-1000017-9234.


\begin{references}
\bibitem{remind} Remind that completely another pattern is observed in case
of short pulses, but for the same values of other parameters
chosen in the text: the $\omega _{2}$ mode is not practically
generated, while the signal $\omega _{1}$ pulse is split into two
temporal modes \cite {sis}. This is because the short pulses are
hardly overlapped in a dense medium and emerge very fast the
interaction region due to the difference in their group
velocities.
\end{references}
\end{document}